# Limited-View and Sparse Photoacoustic Tomography for Neuroimaging with Deep Learning

Steven Guan[1,2*], Amir A. Khan[1], Siddhartha Sikdar[1], and Parag V. Chitnis[1*]

[1]Bioengineering Department, George Mason University., Fairfax, VA USA.
[2]The MITRE Corporation., McLean, VA, 22102. *email: sguan2@gmu.edu; pchitnis@gmu.edu

## Abstract

Photoacoustic tomography (PAT) is a non-ionizing imaging modality capable of acquiring high contrast and resolution images of optical absorption at depths greater than traditional optical imaging techniques. Practical considerations with instrumentation and geometry limit the number of available acoustic sensors and their "view" of the imaging target, which result in image reconstruction artifacts degrading image quality. Iterative reconstruction methods can be used to reduce artifacts but are computationally expensive. In this work, we propose a novel deep learning approach termed pixel-wise deep learning (Pixel-DL) that first employs pixel-wise interpolation governed by the physics of photoacoustic wave propagation and then uses a convolution neural network to reconstruct an image. Simulated photoacoustic data from synthetic, mouse-brain, lung, and fundus vasculature phantoms were used for training and testing. Results demonstrated that Pixel-DL achieved comparable or better performance to iterative methods and consistently outperformed other CNN-based approaches for correcting artifacts. Pixel-DL is a computationally efficient approach that enables for real-time PAT rendering and improved image reconstruction quality for limited-view and sparse PAT.

## Introduction

Neuroimaging in small animals have played an essential role in preclinical research to provide physiological, pathological, and functional insights that are key for understanding and treating neurological diseases. Over the past few decades, there has been significant advances in magnetic resonance imaging (MRI) and optical imaging techniques for structural and functional neuroimaging. For example, MRI can acquire high resolution images of brain structures over large volumes, 3D connectivity and diffusivity information using diffusion tensor imaging, and brain activity using functional MRI [1–3]. However, MRI has poor temporal resolution and cannot be used to study fast hemodynamic mechanisms and responses. Optical imaging techniques can exploit the diverse biological molecules (e.g. hemoglobin, melanin, and lipids) – each possessing different optical properties – present in biological tissues to provide contrast for structural and functional imaging [4–6]. However, strong optical scattering limits the imaging depth of optical techniques to approximately 1-2 mm into the brain [7].

Photoacoustic tomography (PAT) is an emerging non-invasive hybrid technique that has recently seen substantial growth in numerous preclinical biomedical applications and as a powerful clinical diagnostic tool [8–11]. In particular, there is a strong interest in PAT for preclinical structural and functional neuroimaging [12–16]. Given its unique use of light and sound, PAT combines the high contrast and molecular specificity of optical imaging with the high spatial resolution and centimeter-penetration depth of ultrasound imaging [17–19]. PAT has been demonstrated capable of kilohertz volumetric imaging rates, far exceeding the performance of other modalities, which enables new insights into previously obscure biological phenomena [20]. There are diverse contrast agents available such as chemical dyes, fluorescent proteins, and nanoparticles that can be used to further enhance the imaging capabilities of PAT [21,22].

PAT involves irradiating the biological tissue with a short-pulsed laser. Optical absorbers within the tissue are excited by the laser and undergo thermoelastic expansion which results in the generation of acoustic waves [23]. A sensor array surrounding the tissue is then used to detect the acoustic waves, and an image is formed from the measured sensor data. PAT image reconstruction is a well-studied inverse problem that can be solved using analytical solutions, numerical methods (e.g. time reversal), and model-based iterative methods [24–28]. In general, a high-quality image can be reconstructed if the sensor array has a sufficiently large number of sensor elements and completely encloses the tissue. However, building an imaging system with these specifications is often prohibitively expensive, and in many *in vivo* applications such as neuroimaging, the sensor array typically can only partially enclose the tissue [29,30]. These practical limitations result in sparse spatial sampling and limited-view of the photoacoustic waves emanating from the medium. Reconstructing from sub-optimally acquired data causes streaking artifacts in the reconstructed PAT image that inhibits image interpretation and quantification [31].

To address these issues, iterative methods are commonly employed to remove artifacts and improve image quality. These methods use an explicit model of photoacoustic wave propagation and seek to minimize a penalty function that incorporates prior information [32–34]. However, they are computationally expensive due to the need for repeated evaluations of the forward and adjoint operators, and resulting image quality is dependent on the constraints imposed [35,36].

Given the wide success of deep learning in computer vision, there is a strong interest in applying similar methods for tomographic image reconstruction problems [37–39]. Deep learning has the potential to be an effective and computationally efficient alternative to state-of-the-art iterative methods. Having such a method would enable improved image quality, real-time PAT image rendering, and more accurate image interpretation and quantification.

Among the many deep learning approaches for image reconstruction, post-processing reconstruction (Post-DL) is the most widely used and has been demonstrated for improving image reconstruction quality in CT [40,41], MRI [42], and PAT [43–48]. It was shown capable of achieving comparable or better performance than iterative methods for limited-view and sparse PAT image reconstruction [45,49–51]. In Post-DL, an initial inversion is used to reconstruct an image with artifacts from the sensor data. A convolutional neural network (CNN) is then applied as a post-processing step to remove artifacts and improve image quality. The main drawback of Post-DL is that the initial inversion does not properly address the issues of limited-view and sparse sampling, which results in an initial image with artifacts. Image features (e.g. small vessels) that are missing or obscured by artifacts are unlikely to be recovered by the CNN.

Previous works attempted to improve upon Post-DL by removing the need for an initial inversion step [50,52]. One approach termed direct reconstruction (Direct-DL) used a CNN to reconstruct an image directly from the sensor data [52]. The main challenge in using Direct-DL is the need to carefully select parameters (e.g. stride and kernel size) for each convolutional layer in order to transform the sensor data into the desired image dimensions. Changing either the dimensions of the input (e.g. using a different number of sensors) or output would require a new set of convolution parameters and the CNN architecture to be modified. Direct-DL was shown capable of reconstructing an image but underperformed compared to Post-DL. Interestingly, a hybrid approach using a combination of Post-DL and Direct-DL, where an initial inversion and the sensor data are given as inputs to the CNN, was shown to provide an improvement over using Post-DL alone [53,54].

Another approach termed "model-based learning" similarly does not require an initial inversion step and achieves state-of-the-art image reconstruction quality [50,55–57]. This approach is like iterative reconstruction and uses an explicit model of photoacoustic wave propagation for image reconstruction. However, the prior constraints are not handcrafted and instead are learned by a CNN from training data. The improved performance does come at the cost of requiring more time to train the CNN and reconstruct an image [50]. Thus, the choice between model-based learning and direct learned approaches (e.g. Post-DL and Direct-DL) depends on whether the application prioritizes image reconstruction speed or quality.

In this work, we propose a novel approach termed pixel-wise deep learning (Pixel-DL) for limited-view and sparse PAT image reconstruction. Pixel-DL is a direct learned approach that employs pixel-wise interpolation to window relevant information, based on the physics of photoacoustic wave propagation, from the sensor data on a pixel-basis. The pixel-interpolated data is provided as an input to the CNN for image reconstruction. This strategy removes the need for an initial inversion and enables the CNN to utilize more information from the sensor data to reconstruct a higher quality image. The pixel-interpolated data has similar dimensions to the desired output image which simplifies CNN implementation. We compare Pixel-DL to conventional PAT image reconstruction methods (time reversal and iterative reconstruction) and direct learned approaches (Post-DL and a modified implementation of Direct-DL) with *in silico* experiments using several vasculature phantoms for training and testing.

## Methods

### *Photoacoustic Signal Generation*

The photoacoustic signal is generated by irradiating the tissue with a nanosecond laser pulse $\delta(t)$. Light absorbing molecules in the tissue undergo thermoelastic expansion and generate photoacoustic pressure waves [23]. Assuming negligible thermal diffusion and volume expansion during illumination, the initial photoacoustic pressure $x$ can be defined as

$$x(r) = \Gamma(r)A(r) \qquad (1)$$

where $A(r)$ is the spatial absorption function and $\Gamma(r)$ is the Grüneisen coefficient describing the conversion efficiency from heat to pressure [58]. The photoacoustic pressure wave $p(r,t)$ at position $r$ and time $t$ can be modeled as an initial value problem for the wave equation, in which $c$ is the speed of sound [59].

$$(\partial_{tt} - c_0^2\Delta)p(r,t) = 0, \quad p(r,t=0) = x, \quad \partial_t p(r,t=0) = 0 \quad (2)$$

Sensors located along a measurement surface $S_o$ measure a time-dependent signal. The linear operator $\mathcal{M}$ acts on $p(r,t)$ restricted to the boundary of the computational domain $\Omega$ over a finite time $T$ and provides a linear mapping from the initial pressure $x$ to the measured time-dependent signal $y$.

$$y = \mathcal{M}_{p|\partial\Omega\times(0,T)} = Ax \qquad (3)$$

*Photoacoustic Image Reconstruction*

Time reversal is a robust reconstruction method that works well for homogenous and heterogeneous mediums and also for any arbitrary detection geometry [27,28]. A PAT image is formed by running a numerical model of the forward problem backwards in time. This involves transmitting the measured sensor data in a time-reversed order into the medium. Time reversal can reconstruct a high-quality image if the acoustic properties of the medium are known *a priori* and if the sensor array has enough detectors and fully encloses the tissue.

In this work, iterative reconstruction is used to recover the PAT image $x$ from the measured signal $y$ by solving the following optimization problem using the isotropic total variation (TV) constraint

$$x = \underset{x'}{\operatorname{argmin}} \, || \, y - Ax' \, ||^2 + \lambda |x'|_{TV}$$

where the parameter $\lambda > 0$ is a regularization parameter [32,36,60]. The TV constraint is a widely employed regularization functional for reducing noise and preserving edges. Iterative reconstruction with a TV constraint works well in the case of simple numerical or experimental phantoms but often leads to sub-optimal reconstructions for images with more complex structures [43].

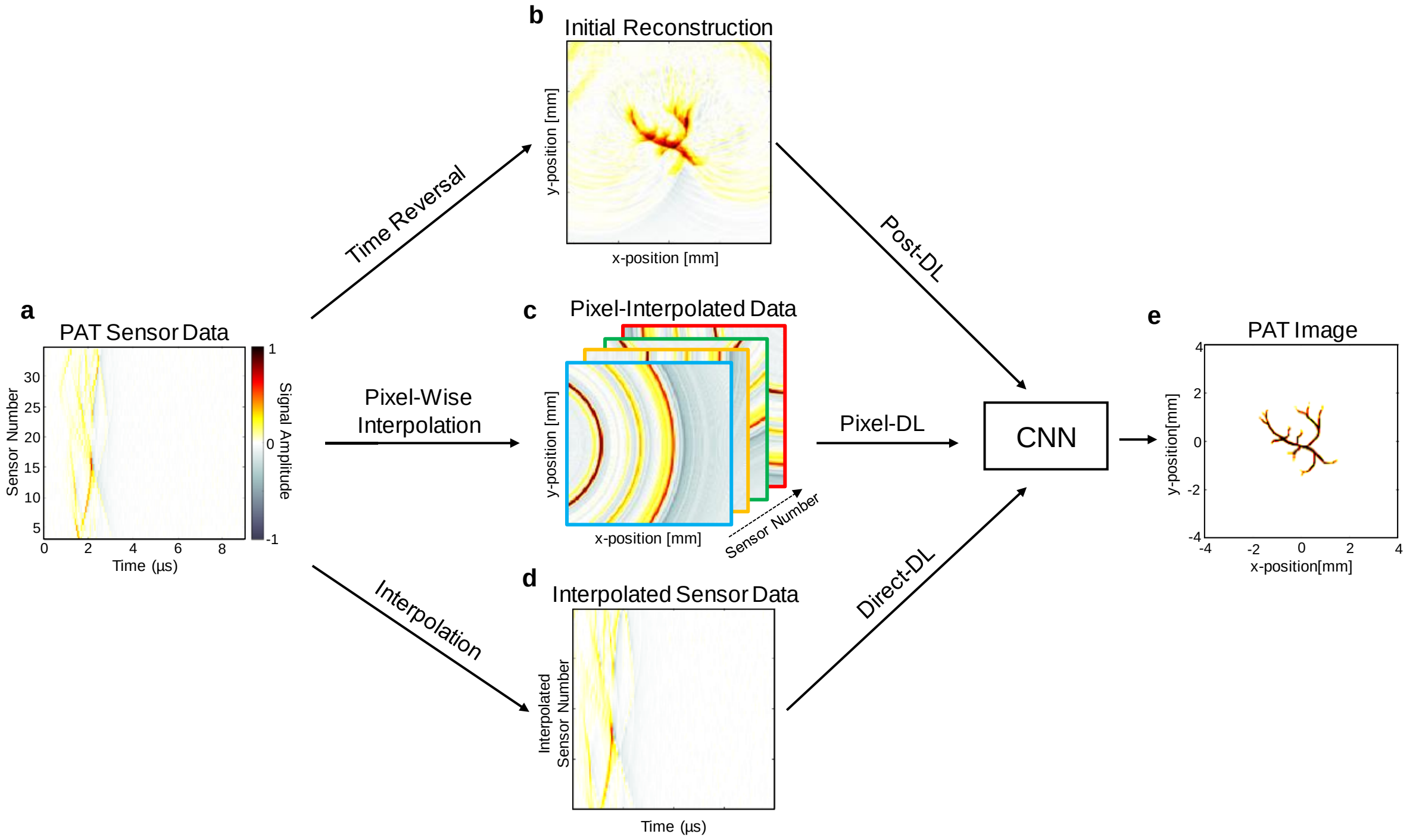

**Fig. 1 | Summary of CNN-based deep learning approaches for PAT image reconstruction**. The primary task is to reconstruct an essentially artifact-free PAT image from the acquired PAT sensor data. **a,** PAT sensor data acquired using a sensor array with 32 sensors and semi-circle limited-view. **b,** Initial image reconstruction with sparse and limited-view artifacts using time reversal for Post-DL. **c,** 3D data array acquired after applying pixel-wise interpolation for Pixel-DL. **d,** Sensor data interpolated to have matching dimensions as the final PAT image for mDirect-DL. **e,** Desired artifact-free PAT image reconstruction from the CNN-based deep learning approaches.

### *Deep Learning*

In this work, three different CNN-based deep learning approaches were used for limited-view and sparse PAT image reconstruction (Fig. 1). These direct learned approaches all began with applying an initial processing step to the PAT sensor data and then recovering the final PAT image using a CNN. The primary difference among these approaches was the processing step used to initially transform the PAT sensor data. In Post-DL, the sensor data was initially reconstructed into an image containing artifacts using time reversal, and the CNN was applied as a post-processing step for artifact removal and image enhancement. In Pixel-DL, pixel-wise interpolation was applied to window relevant information in the sensor data and to map that information into the image space. In the modified Direct-DL implementation (mDirect-DL), a combination of linear interpolation and down sampling was applied so that the interpolated sensor data had the same dimensions as the final PAT image.

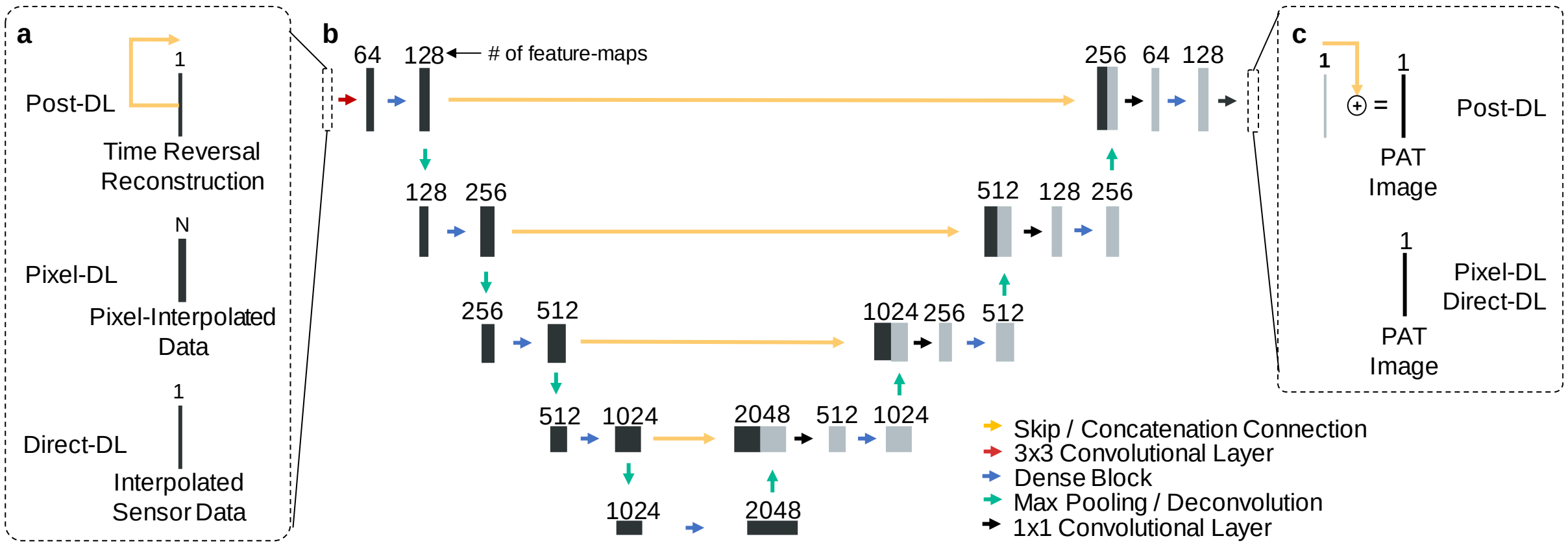

**Fig. 2 | FD-UNet CNN Architecture**. The FD-UNet CNN with hyperparameters of initial growth rate, $k_1 = 16$ and initial feature-maps learned, $f_1 = 128$ is used for PAT image reconstruction. Essentially the same CNN architecture was used for each deep learning approach except for minor modifications. **a,** Inputs into the CNN for each deep learning approach. The Post-DL CNN implementation used residual learning which included a skip connection between the input and final addition operation. The initial Pixel-DL input contains “N” feature-maps corresponding to the number of sensors in the imaging system. **b,** The FD-UNet is comprised of a contracting and expanding path with concatenation connections. **c,** The output of the CNN is the desired PAT image. In Post-DL, residual learning is used to acquire the final PAT image.

### *CNN Architecture: Fully Dense UNet*

After the sensor data was transformed, the final PAT image was recovered using the Fully Dense UNet (FD-UNet) CNN architecture (Fig. 2). The FD-UNet builds upon the UNet, a widely used CNN for biomedical imaging tasks. by incorporating dense connectivity into the contracting and expanding paths of the network [61]. This connectivity pattern enhances information flow between convolutional layers to mitigate learning redundant features and reduce overfitting [62]. The FD-UNet was demonstrated to be superior to the UNet for artifact removal and image enhancement in 2D sparse PAT [47].

### *Pixel-Wise Interpolation*

Pixel-wise interpolation uses a model of photoacoustic wave propagation to map the measured time series pressure in the sensor data to a pixel position within the image reconstruction grid that the signal likely originated from. In this work, we choose to apply pixel-wise interpolation using a linear model of photoacoustic wave propagation since the *in silico* experiments were performed using a homogenous medium (e.g. uniform density and speed of sound). The linear model assumes the acoustic waves are propagating spherically and traveling at a constant speed of sound. Based on these assumptions, the time-of-flight can be easily calculated for a pressure source originating at some position in the medium and traveling to a sensor located on the medium boundary.

Reconstructing an image begins by defining an image reconstruction grid that spans the region of interest in the imaging system (Fig. 3a). The goal of pixel-wise interpolation is to map the time series pressure measurements of each sensor to the defined reconstruction grid on a pixel-basis, which results in a 3D data array with dimensions corresponding to the 2D image space and sensor number (Fig. 3b-c). This is achieved by repeating the following interpolation process for each sensor in the sensor array (Fig. 3d-f). The time-of-flight for a signal originating from each pixel position and traveling to the selected sensor is calculated based on a model of photoacoustic wave propagation. In the case of a linear model, the time-of-flight is proportional to the distance between the selected pixel and sensor (Fig. 3e). Pressure measurements in the sensor data are interpolated onto the reconstruction grid using the calculated time-of-flight for each pixel (Fig. 3f).

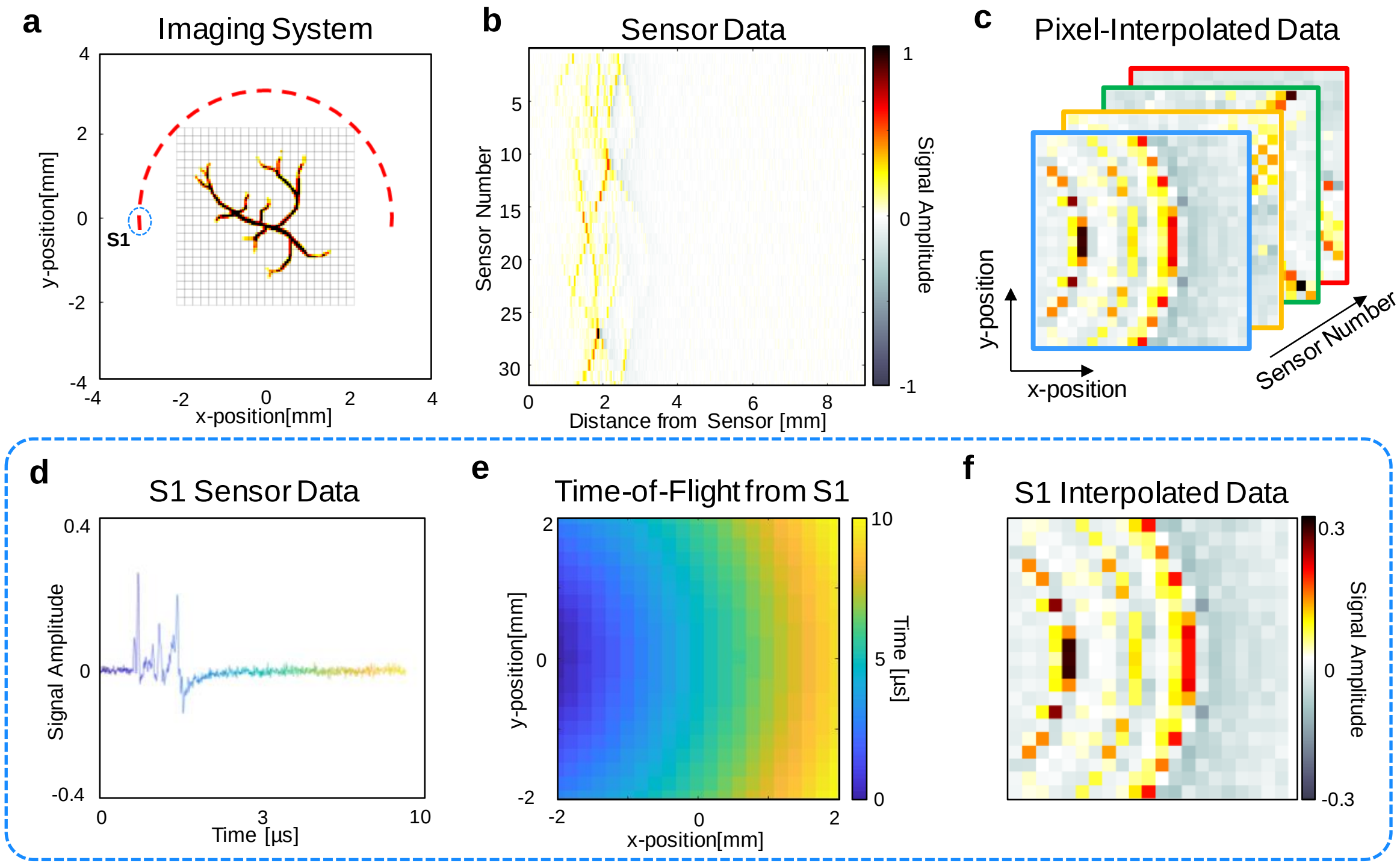

**Fig. 3 | Pixel-Wise Interpolation Process**. **a,** Schematic of the PAT system for imaging the vasculature phantom. The red semi-circle represents the sensor array, and the gray grid represents the defined reconstruction grid. The first sensor (S1) is circled and used as an example for applying pixel-wise interpolation to a single sensor. **b,** The PAT time series pressure sensor data measured by the sensor array. **c,** Resulting pixel-interpolated data after applying pixel-wise interpolation to each sensor based on the reconstruction grid. **d,** Sensor data for S1. Color represents the time at which a pressure measurement was taken and is included to highlight the use of time-of-flight to map the sensor data to the reconstruction grid. **e,** Calculated time-of-flight for a signal originating at each pixel position and traveling to S1. **f,** Pressure measurements are mapped from the S1 sensor data to the reconstruction grid based on the calculate time-of-flight for each pixel.

### *Deep Learning Implementation*

The CNNs were implemented in Python 3.6 with TensorFlow v1.7, an open source library for deep learning [63]. Training and evaluation of the network is performed on a GTX 1080Ti NVIDIA GPU. The CNNs were trained using the Adam optimizer to minimize the mean squared error loss with an initial learning rate of 1e-4 and a batch size of three images for 40 epochs. Training each CNN required approximately one hour to complete. Pairs of training datasets $\{x_i, y_i\}$ were provided to the CNN during training, where $x_i$ represents the input data (e.g. initial time reversal reconstruction, pixel-interpolated sensor data, and interpolated sensor data) and $y_i$ represents the corresponding artifact-free ground truth image. A separate CNN was trained for each CNN-based approached, imaging system configuration, and training dataset.

*Photoacoustic Data for Training and Testing*

Training data were procedurally generated using data augmentation, where new images were created based on a 340x340 pixel-size image of a synthetic vasculature phantom generated in MATLAB (Fig. 3a). First, scaling and rotation was applied to the initial phantom image with a randomly chosen scaling factor (0.5 to 2) and rotation angle (0-359 degrees). Then a 128x128 pixels sub-image was randomly chosen from the transformed image and translated by a random vertical and horizontal shift (0-10 pixels) via zero-padding. Outputs from multiple iterations (up to five) of the data augmentation process are summed together to create a training image. The synthetic vasculature phantom dataset was comprised of 500 training images. Testing data were generated from a 3D micro-CT mouse brain vasculature volume [64] with a size of 260x336x438 pixels. The Frangi vesselness filter was applied to suppress background noise and enhance vessel-like features [65]. A new image was created from the filtered volume by generating a maximum-intensity projection of a randomly chosen 128x128x128 pixel sub-volume. The mouse brain vasculature dataset was comprised of 50 testing images.

The "High-Resolution Fundus Image Database" is a public database that contains 45 fundus images from human subjects that were either healthy, had glaucoma, or had diabetic retinopathy. The images had corresponding vessel segmentation maps created by a group of experts and clinicians within the field of retinal image analysis [66]. The 45 fundus images were split into a separate training set (N=15) and testing set (N=30). The training dataset was procedurally generated using data augmentation based on the images within the training set and was comprised of 500 training images. The testing dataset was comprised of the original 30 images and 20 additional images, generated using data augmentation based on images from the testing set, for a total of 50 testing images.

The "ELCAP Public Lung Image Database" is a public database that contains 50 low-dose whole-lung CT scans obtained within a single breath hold [67]. The whole-lung volumes were split into a training (N=15) and testing set (N=35). Vessel-like structures were segmented from the whole-lung CT volumes using the Frangi vesselness filter [63]. The training dataset was then generated by taking maximum intensity projection images (MIP) of randomly sampled sub-volumes from the filtered volumes in the training set. Data augmentation was also applied to the MIPs to generate a training dataset comprised of 500 training images. With the same procedures, MIPs were taken from the filtered volumes in the testing set to create a testing dataset comprised of 50 images.

In all three cases (mouse-brain vasculature, fundus image database, and ELCAP Lung database), training and testing data were completely segregated. In the latter two experiments, significant variations were present between the training and testing datasets due to patient-to-patient variability and innate differences in vascular morphology between healthy subjects and patients with varying degrees of disease.

A MATLAB toolbox, k-WAVE, was used to simulate photoacoustic data acquisition using an array of acoustic sensors [68]. Photoacoustic simulations in the k-WAVE toolbox are implemented using a pseudospectral approach [69]. Each training and testing image were normalized (values between 0 and 1) and treated as a photoacoustic source distribution on a computation grid of 128x128 pixels. The medium was assumed to be non-absorbing and homogenous with a speed of sound of 1500 m/s and density of 1000 Kg/m$^3$. The sensor array had 16, 32, or 64 sensor elements equally spaced on a semi-circle with a diameter of 120 pixels. The time reversal method in the k-WAVE toolbox was also used for reconstructing an image from the simulated photoacoustic time series data.

Reconstructed images were compared against the ground truth using the peak-signal-to-noise ratio (PSNR) and structural similarity index (SSIM) as metrics for image quality. PSNR provides a global measurement of image quality, while SSIM provides a local measurement that takes into account for similarities in contrast, luminance, and structure [70].

## Results

Conventional PAT image reconstruction techniques (e.g. time reversal and iterative reconstruction) and CNN-based approaches (Post-DL, Pixel-DL, and mDirect-DL) were compared over several *in silico* experiments for reconstruction image quality and reconstruction time. CNN-based approaches were all implemented using the FD-UNet CNN architecture. Reconstructed images were compared to the ground truth image using PSNR and SSIM as quantitative metrics for image reconstruction quality.

### *Mouse Brain Vasculature Experiment*

In the first experiment, the CNNs were trained on the synthetic vasculature phantom dataset and tested on the mouse brain vasculature dataset. Although both datasets contained images of vasculature, they were non-matched meaning there were likely image features (e.g. vessel connectivity patterns) in the testing dataset but not in the training dataset. In addition to evaluating the CNNs' performance, this experiment sought to determine if the CNNs were generalizable when trained on the synthetic vasculature phantom and tested on the mouse brain datasets.

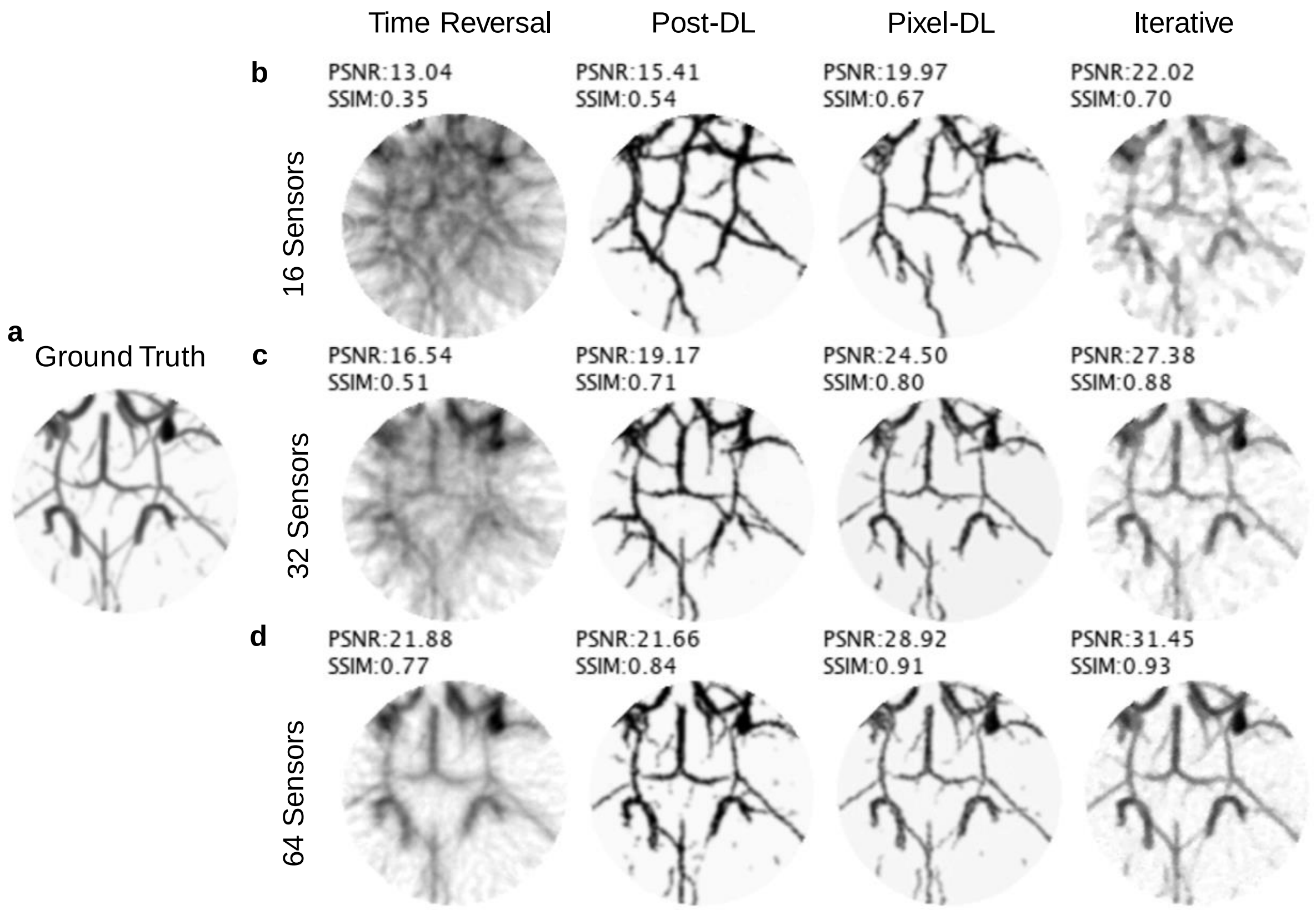

**Fig. 4. | Limited-view and sparse PAT image reconstruction of mouse brain vasculature**. PAT sensor data acquired with a semi-circle limited-view sensor array at varying sparsity levels. **a,** Ground truth image used to simulate PAT sensor data. **b,** PAT reconstructions with 16 sensors. Vessels are difficult to identify in time reversal reconstruction as a result of artifacts. **c,** PAT reconstructions with 32 sensors. Vessels can be clearly seen in CNN-based and iterative reconstructions. **d,** PAT reconstructions with 64 sensors. Larger vessels are identifiable in all reconstructed images.

**Table 1**: Average PSNR and SSIM for Micro-CT Mouse Brain Vasculature Testing Dataset (N = 50 testing images)

| Number of Sensors | Time Reversal | Post-DL | Pixel-DL | Iterative Reconstruction |
|---|---|---|---|---|
| **16** | 13.91±1.12<br>*0.34±0.04* | 17.4±1.24<br>*0.52±0.04* | 21.52±1.36<br>*0.64±0.04* | 22.64±1.4<br>*0.66±0.05* |
| **32** | 17.29±1.20<br>*0.48±0.04* | 21.31±1.10<br>*0.71±0.04* | 25.67±1.29<br>*0.81±0.04* | 26.98±2.11<br>*0.82±0.06* |
| **64** | 22.7±1.06<br>*0.73±0.03* | 24.37±1.25<br>*0.85±0.03* | 29.59±1.42<br>*0.91±0.02* | 30.16±2.70<br>*0.89±0.05* |

For each row, PSNR is shown as normal text on top while SSIM is shown as italicized text on the bottom.

The time reversal reconstructed images had severe artifacts blurring the image and the lowest average PSNR and SSIM for all sparsity levels (Fig. 4 and Table 1). Images reconstructed with iterative or a CNN-based method had fewer artifacts and a higher average PSNR and SSIM. Vessels obscured by artifacts in the time reversal reconstructed images were more visible in the other reconstructed images. As expected, increasing the number of sensors resulted in fewer artifacts and improved image quality for all PAT image reconstruction methods. Pixel-DL consistently had a higher average PSNR and SSIM than Post-DL for all sparsity levels and similar scores to iterative reconstruction.

In the case of sparse sampling (especially with 16 sensors), Post-DL often introduced additional vessels that were not originally in the ground truth image (Fig. 4a-b). This was likely due to the CNN misinterpreting strong artifacts in the input image as real vessels. Pixel-DL exhibited a similar behavior but typically had fewer false additional vessels. This issue was not as prevalent in images reconstructed using the iterative method. However, images reconstructed using iterative reconstruction had an overly smoothed appearance compared to the deep learning-based reconstructed images. This is a pattern commonly observed when using the total variation constraint. Moreover, iterative reconstruct

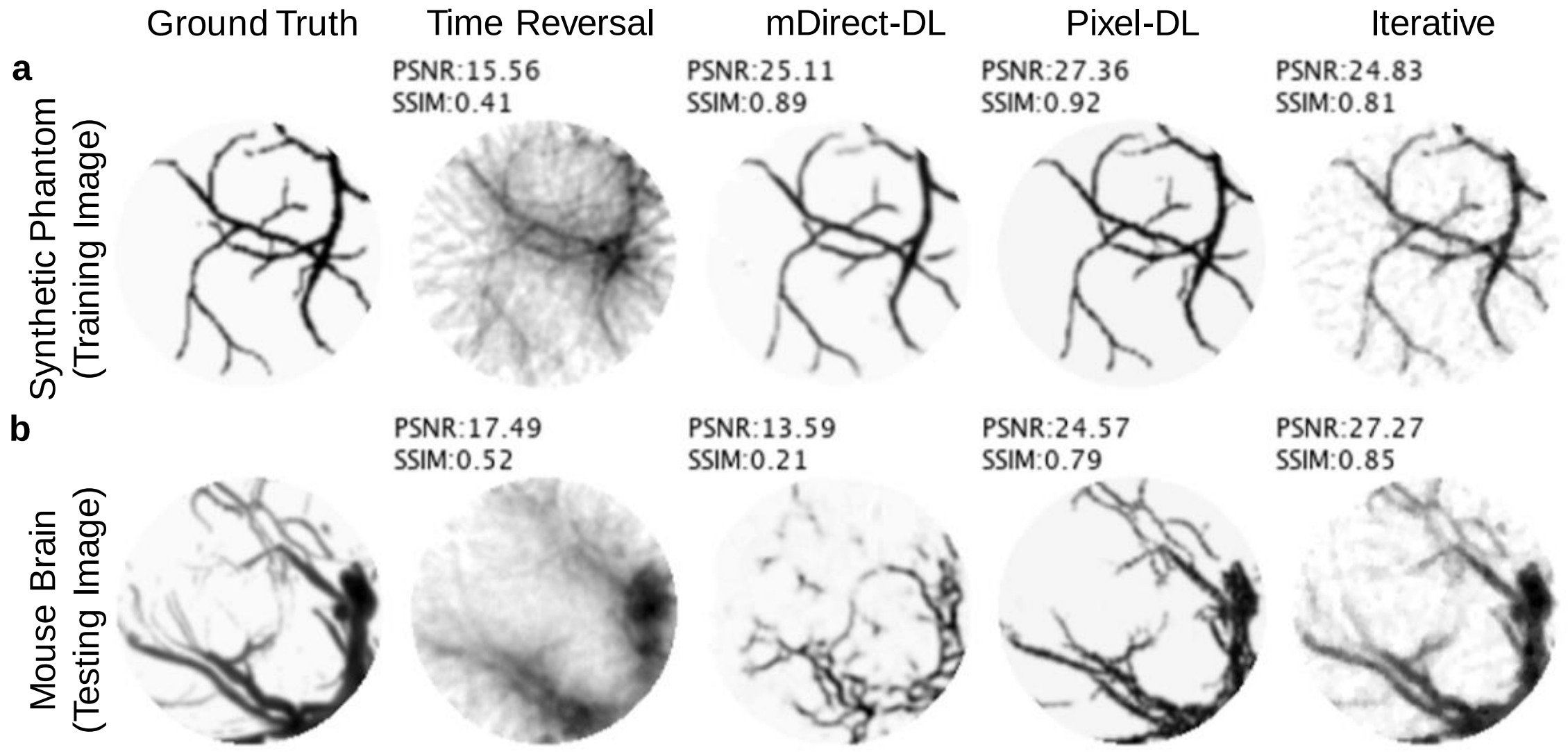

**Fig. 5. | Limited-view and sparse Pixel-DL and mDirect-DL PAT image reconstructions**. PAT sensor data acquired with 32 sensors and a semi-circle view. **a,** CNNs were trained and tested on images of the synthetic vasculature phantom. Both CNN-based approaches successfully reconstructed the example synthetic vasculature phantom image **b,** CNNs were trained on images of the synthetic vasculature phantom but tested on mouse brain vasculature images. mDirect-DL failed to reconstruct the example mouse brain vasculature image and performed worse than time reversal.

Pixel-DL consistently outperformed time reversal in reconstructing images of the synthetic vasculature and mouse brain vasculature (Fig. 5). Interestingly, mDirect-DL only outperformed time reversal in reconstructing the synthetic vasculature images, which were used to train the CNN. The mDirect-DL reconstructed image of mouse brain vasculature resembled the ground truth image but was substantially worse than the time reversal reconstruction. This indicated that the CNN learned a mapping from the PAT-sensor data to the image space but severely overfitted to the training data. During training, the CNNs for Pixel-DL and mDirect-DL converged to a minimum mean squared error, but the Pixel-DL CNN converged to a lower error.

*Lung and Fundus Vasculature Experiment*

In the second experiment, the CNNs were trained and tested on the lung vasculature and fundus vasculature datasets. This experiment represented a scenario in which the training and testing datasets are derived from segregated anatomical image data. There were natural differences between the training and testing datasets since the original images were acquired from healthy patients and those with varying disease severity.

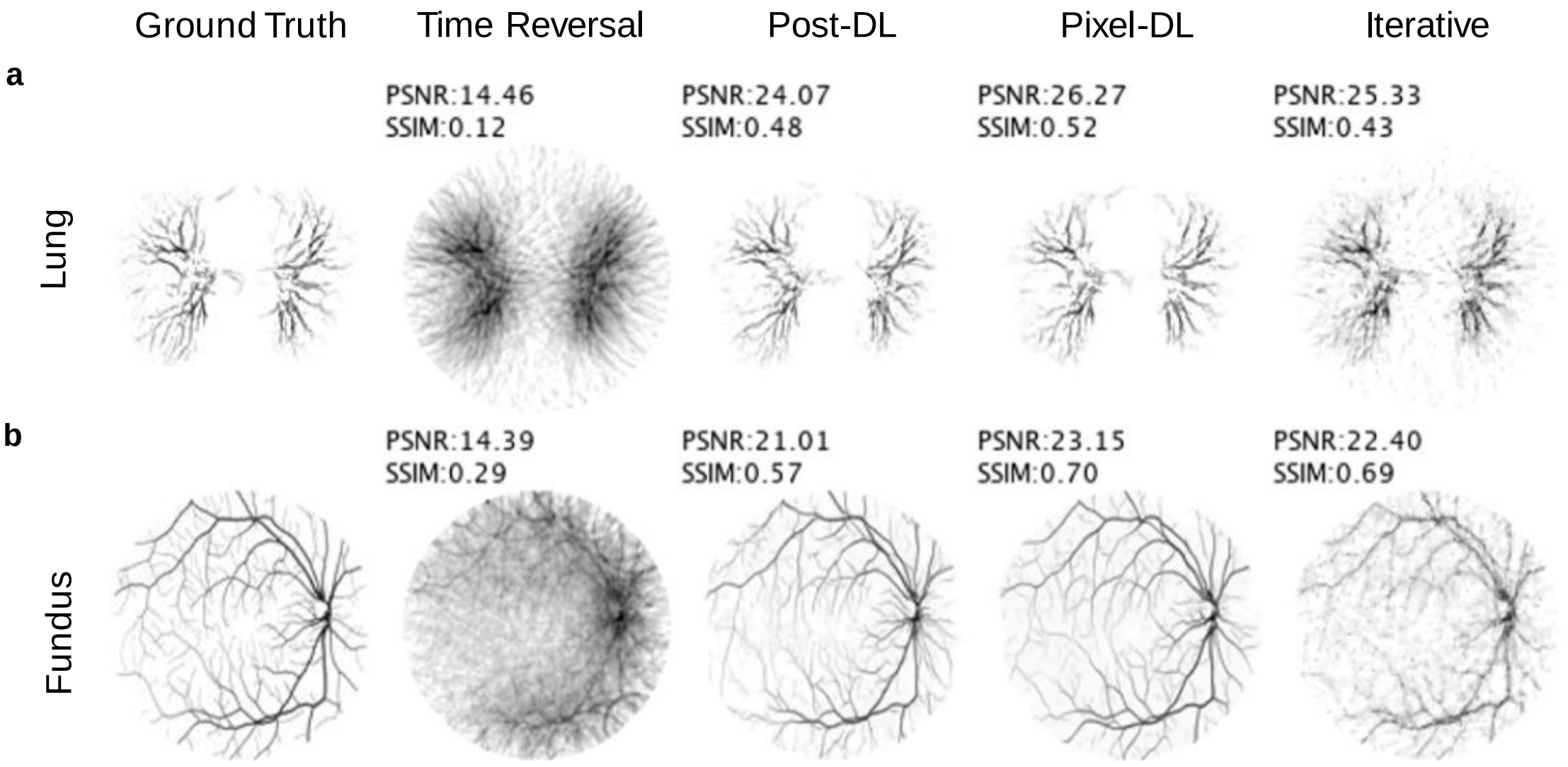

**Fig. 6. | Limited-view and sparse PAT image reconstructions of fundus and lung vasculature**. PAT sensor data acquired with 32 sensors and a semi-circle view. **a,** CNNs were trained and tested on images of lung vasculature **b,** CNNs were trained and tested on images of fundus vasculature. Testing images were derived from a separate set of patients' lung and fundus images than the training images.

**Table 2**: Average PSNR and SSIM for Lung and Fundus Vasculature Testing Dataset (N = 50 testing images)

| | Number of Sensors | Time Reversal | Post-DL | Pixel-DL | Iterative Reconstruction |
|---|---|---|---|---|---|
| **Lung** | 16 | 13.30±1.01<br>*0.09±0.02* | 23.21±1.45<br>*0.35±0.04* | 24.14±1.53<br>*0.43±0.06* | 22.74±1.36<br>*0.29±0.08* |
| | 32 | 15.19±1.13<br>*0.13±0.02* | 25.09±1.67<br>*0.50±0.04* | 26.76±1.83<br>*0.53±0.07* | 27.50±1.98<br>*0.46±0.06* |
| | 64 | 18.82±1.11<br>*0.23±0.05* | 27.14±1.67<br>*0.65±0.04* | 29.98±2.00<br>*0.69±0.11* | 33.67±1.92<br>*0.62±0.07* |
| **Fundus** | 16 | 12.26±1.10<br>*0.19±0.02* | 20.00±1.52<br>*0.42±0.06* | 20.78±1.61<br>*0.52±0.08* | 20.77±1.07<br>*0.50±0.04* |
| | 32 | 14.07±1.38<br>*0.26±0.03* | 21.57±1.60<br>*0.59±0.04* | 23.40±1.40<br>*0.67±0.05* | 23.37±1.06<br>*0.68±0.04* |
| | 64 | 18.08±1.40<br>*0.45±0.05* | 24.16±1.56<br>*0.75±0.03* | 26.23±1.35<br>*0.81±0.05* | 28.07±1.10<br>*0.85±0.06* |

For each row, PSNR is shown as normal text on top while SSIM is shown as italicized text on the bottom.

As expected, the time reversal reconstructed images of lung and fundus vasculature had the most artifacts and the lowest average PSNR and SSIM for all sparsity levels (Fig. 6 and Table 2). Images reconstructed with a CNN-based method or iterative reconstruction resulted in fewer artifacts and a higher average PSNR and SSIM. Pixel-DL consistently outperformed Post-DL for both vasculature phantoms for all sparsity levels. Comparable to iterative reconstruction, Pixel-DL had similar performance for the fundus vasculature and outperformed it for the lung vasculature dataset. For images reconstructed from PAT sensor data acquired using 16 sensors, Pixel-DL reconstructed images appeared sharper and were qualitatively superior compared to iteratively reconstructed images despite having similar SSIM and PSNR values.

### *Image Reconstruction Times*

The average reconstruction time reported for each method are for reconstructing a single image from the PAT sensor data. Time reversal is a robust and computationally inexpensive reconstruction method (~2.57 seconds per image). Iterative reconstruction removed most artifacts and improved image quality but had a much longer average reconstruction time (~491.21 seconds per image). Pixel-DL reconstructed images with similar quality to iterative reconstruction and was faster by over a factor of 1000 (~7.9 milliseconds per image). Average reconstruction time for Post-DL is dependent on the initial inversion used since the computational cost of a forward pass through a CNN is essentially negligible. Since time reversal was used as the initial inversion, Post-DL had a longer average reconstruct time than Pixel-DL (~2.58 seconds per image).

## Discussion

In this work, we propose a novel deep learning approach termed Pixel-DL for limited-view and sparse PAT image reconstruction. We performed *in silico* experiments using training and testing data derived from multiple vasculature phantoms to compare Pixel-DL with conventional PAT image reconstruction methods (time reversal and iterative reconstruction) and direct learned approaches (Post-DL and mDirect-DL). Results showed that Pixel-DL consistently outperformed time reversal, Post-DL, and mDirect-DL for all experiments. Pixel-DL was able to generalize well evidenced by its comparable performance to iterative reconstruction for the mouse brain vasculature phantom despite having only trained on images generated from a synthetic vasculature phantom with data augmentation. Having a more varied training dataset may further improve CNN generalization and performance. When the training and testing data were derived from segregated anatomical data, Pixel-DL had similar performance to iterative reconstruction for the fundus vasculature phantom and outperformed it for the lung vasculature phantom. The total variation constraint used for iterative reconstruction was likely suboptimal for reconstructing lung vasculature images since the lung vessels were small and closely grouped.

### *Comparison between Deep Learning Approaches*

The CNN architecture and hyperparameters used for all deep learning approaches implemented were essentially the same. Thus, discrepancies in performance between the approaches were primarily due to their respective inputs into the CNN (Fig. 4). In Post-DL, the input was an image initially reconstructed from the sensor data using time reversal. The input and output to the CNN are both conveniently images of the same dimensions. This removed the need for the CNN to learn the physics required to map the sensor data into the image space. However, the initial inversion did not properly address the issues of limited-view and sparse sampling which resulted in an initial image with artifacts. Moreover, the CNN no longer had access to the sensor data and was only able to use information contained in the image to remove artifacts. There was likely useful information in the sensor data for more accurately reconstructing the PAT image, which was ignored in this approach.

In Pixel-DL, the initial inversion is replaced with pixel-wise interpolation, which similarly provides a mapping from the sensor data to image space. Relevant sensor data is windowed on a pixel-basis using a linear model of acoustic wave propagation. This enables the CNN to have a richer information source to reconstruct higher quality images. Furthermore, there is no initial inversion introducing artifacts; thus, the CNN does not have an additional task of learning to remove those artifacts.

mDirect-DL similarly did not require an initial inversion and instead used the full sensor data as an input to the CNN to reconstruct an image. The potential advantage of mDirect-DL is that the CNN had full access to the information available in the sensor data to reconstruct a high-quality image. However, reconstructing directly from the sensor data was also a more difficult task because the CNN needed to additionally learn a mapping from the sensor data into the image space. Results showed that the CNN had difficulty in learning a generalizable mapping and overfitted to the training data (Fig. 5). The FD-UNet was likely not an optimal architecture for this task since it was designed assuming the input was an image. A different neural network architecture for a multidimensional time-series input would be better suited.

A limitation of Post-DL and Pixel-DL for sparse and limited-view PAT is that the reconstructed image could have additional vessels that are not in the ground truth image. This can be problematic depending on the requirements of the application. Large vessels and structures are often reliably reconstructed in the image, but some small vessels could be false additions. This limitation primarily occurred at the sparsest sampling level and could be addressed by increasing the number of sensors used for imaging. The loss function could also be modified to penalize the CNN for reconstructing false additional vessels, but this could lead to the CNN to preferentially not reconstruct small vessels. Alternatively, a model-based learning approach could be used for better image quality if computational cost is not a limitation.

### *Deep Learning for In Vivo Imaging*

A key challenge in applying deep learning for *in vivo* PAT image reconstruction is that a large training dataset is required for the CNN to learn and be able to remove artifacts and improve image quality. The training data can be acquired experimentally using a PAT imaging system that has a sufficient number of sensors and full-view of the imaging target. However, this process is often infeasible because it is prohibitively expensive, time-consuming, and needs to be repeated when the imaging system configuration or imaging target is changed. Alternatively, synthetic training data can be generated using numerical phantoms or images from other modalities. In combination with data augmentation techniques, this approach enables for arbitrarily large synthetic training datasets to be created. However, CNN image reconstruction quality is largely dependent on the degree to which the simulations used to generate the training data matches actual experimental conditions. Properly matching the simulation is a non-trivial task that necessitates the PAT imaging system to be well-characterized and understood. Some factors to be considered when creating the simulations include: sensor properties (e.g. aperture size, sensitivity, and directivity), sensor configuration, laser illumination, and medium heterogeneities. Generally, it is preferable to closely match the simulation to the experimental conditions, but post-processing (e.g. filtering and denoising) can also be applied to the experimental data. It is beyond the scope of this work to discuss the impact of each factor in detail, but the issue of medium heterogeneities, specifically for speed of sound, is examined.

In this work, Pixel-DL was applied using a linear model of acoustic wave propagation that assumes the acoustic waves propagate spherically and travel at a constant speed of sound throughout the medium. Although this model was sufficient for the case of a homogenous medium, a different model would be needed if the medium was heterogeneous (e.g. speed of sound and density) such as for *in vivo* imaging. Naively reconstructing with these assumptions for heterogeneous mediums would result in additional artifacts that degrade image quality and potentially impact CNN performance. The severity of the artifacts would depend on the degree of mismatch between the heterogeneity and assumed value. If the distribution of the heterogeneities or acoustically reflective surfaces is known then they can be accounted for during the time-of-flight calculations when applying pixel-interpolation. However, if it is not known then the CNN should be trained with training data containing examples of heterogeneous mediums similar to what would be anticipated during image reconstruction. This would enable the CNN to learn to compensate for potential artifacts due to applying pixel interpolation with a linear model of acoustic wave propagation when the medium is actually not homogeneous.

*Deep Learning for Fast Image Reconstruction*

The proposed Pixel-DL approach can be used as a computationally efficient method for improving PAT image quality under limited-view and sparse sampling conditions. It can be readily applied to a wide variety of PAT imaging applications and configurations. Pixel-DL enables for the development of more efficient data acquisition approaches. For example, PAT imaging systems can be built with fewer sensors without sacrificing image quality, which would allow for the technology to be more affordable. Pixel-DL achieved similar or better performance and was faster than iterative reconstruction by over a factor of a 1000. It would allow for real-time PAT image rendering which would provide valuable feedback during image acquisition.

In this work we have demonstrated *in silico* the feasibility of Pixel-DL for PAT imaging of vasculature-like targets. This approach can also be readily applied to ultrasound imaging. Image reconstruction for PAT and ultrasound imaging both largely rely on time-of-flight calculations to determine where the signal originated. Therefore, a similar linear model of acoustic wave propagation can be used to readily apply Pixel-DL for ultrasound image reconstruction problems. Pixel-DL can also be adapted to other imaging modalities if a model mapping the sensor data to the image space is available.

## Author Contribution

Steven Guan was responsible for the data analysis, wrote the main manuscript, and prepared the figures. Amir A. Khan and Siddhartha Sikdar were advisors that have reviewed and edited the manuscript in addition to playing a critical role in defining the research plan. Parag V. Chitnis was a key contributor in planning and completing the research plan and have also reviewed and made substantial edits to the manuscript.

## Data Availability

The datasets generated during and/or analyzed during the current study are available from the corresponding author on reasonable request.

## Competing Interests

The authors declare no competing interests.